\documentclass{aastex}          
\usepackage{spr-astr-addons}    
\begin{document}
\title{A statistical study of the subsurface structure and eruptivity
of solar active regions}
\shorttitle{}
\shortauthors{Lin}
\author{C.-H. Lin}
\affil{Graduate Institute of Space Sciences,
    National Central University,
    Chung-Li, Taoyuan, Taiwan}
\email{chlin@jupiter.ss.ncu.edu.tw} 
\begin{abstract}
A statistical study of 77 solar active regions (ARs) is conducted to
investigate the existence of identifiable correlations between
the subsurface structural disturbances and
the activity level of the active regions.
The disturbances examined in this study are
$<|\delta \Gamma_1/\Gamma_1|>$,
$<|\delta c^2/c^2|>$, 
and
$<|\delta c^2/c^2-\delta \Gamma_1/\Gamma_1|>$.
where $\Gamma_1$ and $c$ are the thermodynamic properties of 
first adiabatic index and
sound speed modified by magnetic field, respectively.
The averages are over three depth layers:
$0.975-0.98 R_\odot$, $0.98-0.99 R_\odot$ and $0.99-0.995 R_\odot$
to represent the structural disturbances in that layer.
The level of the surface magnetic activity is measured by the
Magnetic Activity Index (MAI) of active region and
the relative and absolute MAI differences (rdMAI and dMAI)
between the active and quiet regions.
The eruptivity of each active region is quantified by its Flare Index,
total number of coronal mass ejections (CMEs),
and total kinetic energy of the CMEs.
The existence and level of the correlations are evaluated by
scatter plots and correlation coefficients.
No definitive correlation can be claimed from the results.
While a weak positive trend is visible between dMAI and 
$<|\delta \Gamma_1/\Gamma_1|>$ and
$<|\delta c^2/c^2|>$ in the layer $0.975-0.98 R_\odot$,
their correlation levels, being approximately 0.6,
are not sufficiently high to justify the correlation.
Some subsurface disturbances are seen to increase with eruptivity indices
among ARs with high eruptivity.
The statistical significance of such trend, however, cannot be ascertained
due to the small number of very eruptive ARs in our sample.
\end{abstract}
\section{Introduction} \label{sec:intro}
Thanks to the increasing impact of space weather on modern society,
many studies have been conducted to examine
the relationship
between the observed photosphere magnetic field
and the production of solar flares
\citep[e.g.,][to name a few]{MGI2005ApJ631,GR2007ApJ661L,LB2007ApJ656,Schrijver2007ApJ662L,LGR2007ApJ,Song_etal2009SoPh,PCW2010ApJ,Ahmet_etal2013SoPh}.
The contributions of the subsurface flow dynamics to 
the flare production have also been investigated
\citep[e.g.,][]{Komm_etal_2004ApJ,Reinard_etal2010ApJ,Komm_etal_2011SoPh}.
Despite different results reported from different studies,
it is generally agreed that
the complexity of magnetic field 
(or the deviation from a potential field) and 
the twisting of the foot points of field lines
increase magnetic energy and stresses in the field,
resulting in a more favorable environment for solar eruptions,
and that the solar eruptions remove the excess magnetic energy
and stresses from the field,
returning it to a lower energy, more stable state.

In contrast,
the relationship between 
the productivity of solar eruptions
and thermal properties of the subsurface structure is largely unknown. 
It is uncertain whether
the subsurface thermal structures
can be related to solar eruptions through
some physical mechanisms.
Because of high gas-to-magnetic pressure ratio below the solar surface,
the relationship, even if exists, can be expected to be very weak,
and the average thermal structure should not show appreciable changes
over the timescale of one eruption.
Therefore,
the relationship is likely to be detectable only among the active regions with
sufficiently high eruptivity level and/or significant subsurface disturbances.
The objective of this work is
to investigate whether
such relationship can be detected with current level of observational
and technological capability. 
The strategy is to conduct a statistical study on the relationships between
the disturbances of subsurface structural properties and
the levels of both the coronal eruptivity and surface magnetic activity of
the solar active regions (ARs).
The results can shed lights on
the physics involved in
the interactions between gas and the magnetic field
and the connection from below to above the solar surface.

The relationship between the subsurface thermal anomalies and 
surface magnetic activity has been examined by
\citet{Bogartetal2008SoPh} and \citet{bbb2013SoPh},
and a positive correlation was claimed by both studies.
\citet{Bogartetal2008SoPh} also examined
the relationship between the subsurface thermal anomalies
and total flare activity,
but found no correlation between the two. 
However, solar flares are not the only eruptive phenomenon in the corona.
The reported un-correlation with flare activity therefore
does not completely rule out the possibility of a correlation
with the eruptivity of active region and/or
the productivity of other types of strong eruptions.
Here we considered the contributions from 
both flares and coronal mass ejections (CMEs),
two largest eruptive phenomena in the corona,
in the assessment of the eruptivity of a region,
and then conducted 
a statistical analysis of 77 active regions
to examine the relationship
between each available subsurface thermal properties
and different indices that characterize the 
surface magnetic activity
and coronal eruptivity.
The subsurface structural differences of these regions
are a subset of the inversion results of \citet{bbb2013SoPh}.

The details of the data source,
the definition of different indices
and the analysis procedures
are described in Sec.~\ref{sec:analysis},
the results are discussed in Sec.~\ref{sec:results},
and
a summary of the results is given in Sec.~\ref{sec:sum}.

\section{Data \& Analysis} \label{sec:analysis}
\subsection{Disturbances of the subsurface structure}
The disturbed thermal properties examined in this study are
$<|\delta c^2/c^2|>$, $<|\delta \Gamma_1/\Gamma_1|>$ and
$<|\delta c^2/c^2 - \delta \Gamma_1/\Gamma_1|>$, in which
$c$ is the thermodynamic property sound speed
modified by the existence of magnetic field,
and $\Gamma_1$ is the adiabatic index defined as
$(\partial \ln P/\partial \ln \rho)_s$, 
where $P$, $\rho$  and $s$ are gas pressure, density and entropy.
We emphasize that $c$ is {\em not} the travel speed of wave
but a thermodynamic property defined as $\Gamma_1 P/\rho$.
In a region free of magnetic field and away from ionization zones,
$P$ is simply the pressure of gas, and
$\delta c^2/c^2$ is a direct representation of the difference in temperature.
However,
because $\delta c^2/c^2$ used in this study is the inversion result
of solar active regions, 
which contain strong magnetic fields,
$P$ is not simply gas pressure,
and $\delta c^2/c^2$ in this case 
cannot directly reflect the temperature difference.
$\delta \Gamma_1/\Gamma_1$ mainly results from a difference in the ionization degree
of gas or the equations of state. 
Thus, the quantity $\delta c^2/c^2 - \delta \Gamma_1/\Gamma_1$
can represent the part of the structural difference that is {\em not}
due to the change of the ionization degree or the equations of state.

$\delta c^2/c^2$ and $\delta \Gamma_1/\Gamma_1$ were provided by
\citet[][private communication]{bbb2013SoPh}, 
who applied ring-diagram analysis \citep{Hill1988ApJ}
and local helioseismic inversion to obtain 
the differences between ARs
and their co-latitude reference quiet-Sun regions (QSs).
The data used in their analysis were the Dopplergrams from 
the Michelson Doppler Imager (MDI) instrument on-board
the Solar and Heliospheric Observatory (SoHO).
The results are the depth profiles of the relative differences
averaged over a $15^{\circ}$ patch in space and $4-7$ days in time.
Since the difference between $\delta c^2/c^2$ and $\delta \Gamma_1/\Gamma_1$
is one of the quantities examined in present work,
only those AR-QS pairs that have
both $\delta c^2/c^2$ and $\delta \Gamma_1/\Gamma_1$ inversion results
were included in this study.
There are total 77 pairs.
This collection of ARs covers time period from 1996 July to 2010 November.
Noticing that $\delta c^2/c^2-\delta \Gamma_1/\Gamma_1$ is generally
large only in $0.98 - 0.99 R_\odot$ and small elsewhere,
I divided the region where the inversion results are most reliable
into three layers:
$0.975 - 0.98 R_\odot$, $0.98 - 0.99 R_\odot$ and $0.99 - 0.995 R_\odot$.
The absolute values of the relative differences
were averaged over these three depth ranges to represent
the average structural disturbances in these layers.

\subsection{Indices of surface magnetic activity levels} 
The level of surface magnetic activity of a region
is characterized by
the Magnetic Activity Index \citep[MAI; ][]{Basuetal2004ApJ},
which is defined as the total strong unsigned magnetic flux within
the area of inversion ($15^{\circ}$ patch) averaged over the tracking period
($4-7$days).
In other words,
the absolute values (in the unit of Gauss) of the strong fields
are integrated over the strong-field area within the region of analysis,
and averaged over the time interval of the analysis.
The MAI of each region used in this work
was provided by
\citet[][private communication]{bbb2013SoPh},
who computed the values using the MDI magnetograms.
\citet{Bogartetal2008SoPh} had reported positive correlation
between the inversion results and MAI of AR (MAI$_{\rm AR}$),
and \citet{bbb2013SoPh} also claimed a positive envelope between
$\delta c^2/c^2$ and the absolute difference of MAI between AR and QS.
However, 
the inversion results are
the {\em relative} structural differences between AR and QS region.
If the magnitudes of MAI of QS regions 
are negligible compared with
those of their pairing ARs,
MAI$_{\rm AR}$ would be equivalent to MAI$_{\rm AR}$-MAI$_{\rm QS}$, 
and would be appropriate to represent the difference
in the magnetic activity between AR and QS.
However, the MAI of several QS regions in our data set are more than 10\% of 
the MAI of their pairing ARs.
Therefore,
in this study,
we examined the relationships between
the subsurface relative differences
and three variations of the magnetic activity indices:
MAI$_{\rm AR}$,
dMAI$\equiv$MAI$_{\rm AR}-$MAI$_{\rm QS}$
and 
rdMAI$\equiv$(MAI$_{\rm AR}-$MAI$_{\rm QS}$)/MAI$_{\rm QS}$.

\subsection{Indices of coronal eruptivity levels}
An ideal index to represent the level of eruptivity would be one that
appropriately incorporate contributions of all types of eruptions.
However, 
there is no generally accepted method to 
combine different types of eruptions.
To avoid subjective bias in making the combination, 
the productivity of different types of eruptions was measured separately.
The specific eruptions considered in this study were
solar flares and coronal mass ejections,
which are the two major eruptive phenomena in the corona.
The level of eruptivity of each active region
was thus gauged by its productivity of these two types of eruptions.
Three indices, Flare Index (FI), total number of associated CMEs (Ncme)
and total kinetic energy of the CMEs (KEsum)
were derived to assess the productivity.

To derive the indices,
a data base
of the flare and CME events associated with each region during its visible time
on the solar surface
was constructed.
The visible time was determined by checking the dates of
the appearance and disappearance
of each AR in
SolarMonitor \footnote{{\tt http://www.solarmonitor.org}}.
The CME events and their source regions were identified
by examining images from different sources 
including
EIT 
\citep[Extreme-Ultraviolet Imaging Telescope; ][]{EIT1995SoPh}
daily movies 
\footnote{{\tt http://lasco-www.nrl.navy.mil/daily\_mvi}},
SOHO LASCO
\citep[Large Angle and Spectrometric Coronagraph; ][]{LASCO1995SoPh}
CME catalog 
\footnote{{\tt http://cdaw.gsfc.nasa.gov/CME\_list/}},
STEREO/SECCHI data
(Solar TErrestrial RElations Observatory/Sun
Earth Connection Coronal and Heliospheric Investigation;
\citet{STEREO2008SSRv}),
and
STEREO COR1 CME catalog 
\footnote{{\tt http://cor1.gsfc.nasa.gov/catalog/}}.
The SOHO LASCO CME catalog was also used for the information of
estimated kinetic energy of most of the CMEs.
The flare information was based on the
NOAA/USAF Active Region Summary
\footnote{{\tt http://www.swpc.noaa.gov/Data/index.html\#reports}}.

Flare index (FI)
is a quantity to quantify the daily flare activity over 24 hours
per day \citep{FI_Antalova1996},
and is defined as
\begin{equation} \label{eqn:FI}
FI=\frac{100 \times S^{X} + 10 \times S^{M} + 1 \times S^{C} 
    + 0.1 \times S^{B}}{T},
\end{equation}
where $T$ is the time interval (measured in days),
and $S^{(i)}$ is the sum of the significances
of the peak flux ($W/m^2$) of flare class $i$ as measured by
GOES
(Geostationary Operational Environmental Satellite;
\citet{GOES1994SoPh})
over the interval $T$.

Ncme is the total number of CMEs associated with an AR
during its entire visible time.
Each event, irrespective of its strength, is equally counted.
There are several fainter ejections that were seen
in EIT and STEREO COR1 but were not listed in the catalog.
The number of these unlisted events is considered as
the uncertainty of Ncme.

KEsum is the sum of the kinetic energy (KE) of the CMEs
divided by 1E+30 erg,
which is the average magnitude of the KE of an CME.
The scaling was to reduce the magnitude of KEsum to avoid numerical problems.
The LASCO CME catalog provides an estimated representative kinetic energy
for many events.
However,
there are also many events that are listed but do not have an estimated KE
in the catalog. 
These are often relatively faint events or
events propagating in a direction that
does not allow the measurements of the linear speed or mass from
the view point of LASCO.
The KE of such event was given a value of 1.E+29 erg.
Event that was {\em not} detected by LASCO C2 but was seen in
EIT and STEREO COR1
was given a value of 1.E+28 erg as KE because
such event is usually fainter than the average ones.
The sum of all the unlisted KEs
is considered as the uncertainty of KEsum.

In short,
there are three indicators for the subsurface structural disturbances
in three depth ranges:
$<|\delta c^2/c^2|>$, $<|\delta \Gamma_1/\Gamma_1|>$ and
$<|\delta c^2/c^2 - \delta \Gamma_1/\Gamma_1|>$
in $0.975-0.98 R_\odot$, $0.98-0.99 R_\odot$ and $0.99-0.995 R_\odot$,
three indices for the magnetic activity on the surface 
(rdMAI, dMAI and MAI$_{\rm AR}$),
and three indices for the eruptivity in the corona
(FI, Ncme and KEsum).

\subsection{Analysis method}
To simplify the notations for the rest of the paper,
Y(i, j) is used to represent the three 
averaged subsurface structural differences:
Y(0, j)  to Y(2, j) denote
$<|\delta \Gamma_1/\Gamma_1|>$, $<|\delta c^2/c^2|>$
and $<|\delta c^2/c^2-\delta \Gamma_1/\Gamma_1|>$, respectively, 
and Y(i, 0) to Y(i, 2) represent
the three depth ranges
$0.975-0.98 R_\odot$, $0.98-0.99 R_\odot$
and $0.99-0.995 R_\odot$, respectively.
The six activity/eruptivity indices are denoted by X(k):
X(0) to X(5) stand for rdMAI, dMAI, MAI$_{\rm AR}$, 
FI, Ncme and KEsum, respectively.
The plan was to go through all possible combinations of Y(i,j) versus X(k)
to see if any correlation can be identified between the subsurface disturbances
and the activity/eruptivity above the surface.
For each combination,  
a scatter plot was generated for visual inspection,
and 
the level of correlation was assessed by the correlation coefficient
(CC) \citep{Barlow}:
\begin{eqnarray} \label{eqn:corr}
{\rm CC} &=& \frac{{\rm cov}(x, y)}{\sigma_x \sigma_y} \\
          &=& \frac{\overline{x y}-\overline{x} \; \; \overline{y}}{\sigma_x \sigma_y}
\end{eqnarray}
where $x$ and $y$ are two linearly related variables,
$\sigma_x$ and $\sigma_y$ are their respective standard deviations,
and $\overline{x}$, $\overline{y}$ and $\overline{x y}$ are the means of
$x$, $y$ and $x y$, respectively.
However, the correlation coefficient 
should not be used as a confirmation or rejection of
the existence of a correlation 
\citep{NumRecC}
because
there is no universal way to evaluate the significance of
the value of the correlation coefficient in different situations. 
Hence,
in this study,
it is only used as a reference to quantify the level of correlation
after a linear trend is visually identified in the scatter plots,
and CC=$0.6$ was chosen as the threshold for a correlation to be significant.

\section{Results and discussion} \label{sec:results}

\begin{figure*}
\small
\includegraphics[angle=90,width=17cm]{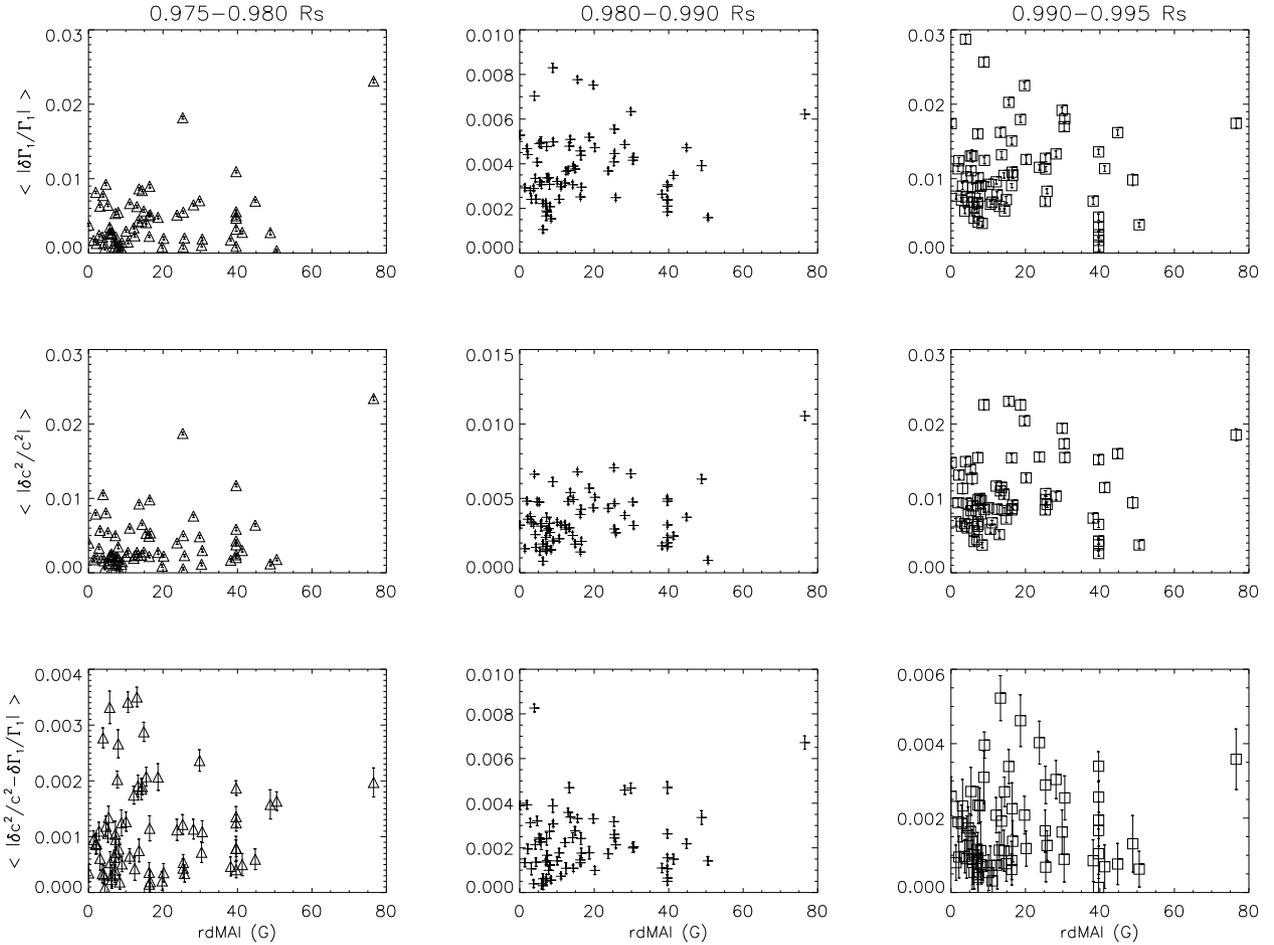}
\caption{
The scatter plots of
the subsurface structural disturbances in different depths vs. rdMAI.
Different structural differences are placed in different rows:
$<|\delta \Gamma_1/\Gamma_1|>$ (top), $<|\delta c^2/c^2|>$ (middle),
$<|\delta c^2/c^2-\delta \Gamma_1/\Gamma_1|>$ (bottom),
as labeled on the left of each row.
Results of different layers from deep to shallow are separated into columns
from left to right,
as indicated on the top of each column.
Each point corresponds to one active region, 
and is located according to Y(i,j) and X(k) of that active region.
}
\label{fig:rdMAI}
\end{figure*}

\begin{figure*}
\small
\includegraphics[angle=90,width=17cm]{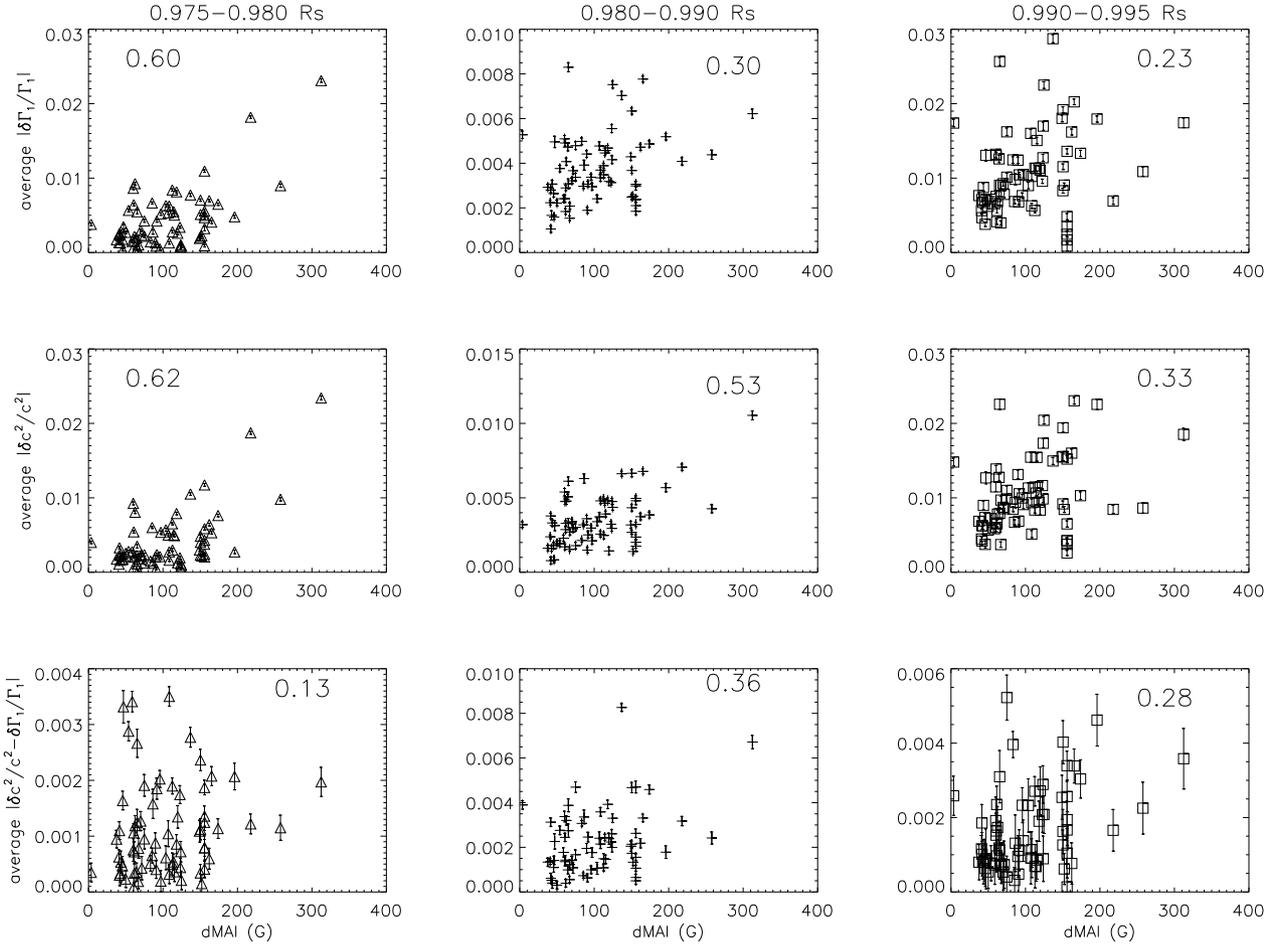}
\caption{
The scatter plots of
the subsurface structural disturbances in different depths vs. dMAI.
The arrangement of the panels and symbols
is the same as in Fig.~\ref{fig:rdMAI}.
}
\label{fig:dMAI}
\end{figure*}

\begin{figure*}
\small
\includegraphics[angle=90,width=17cm]{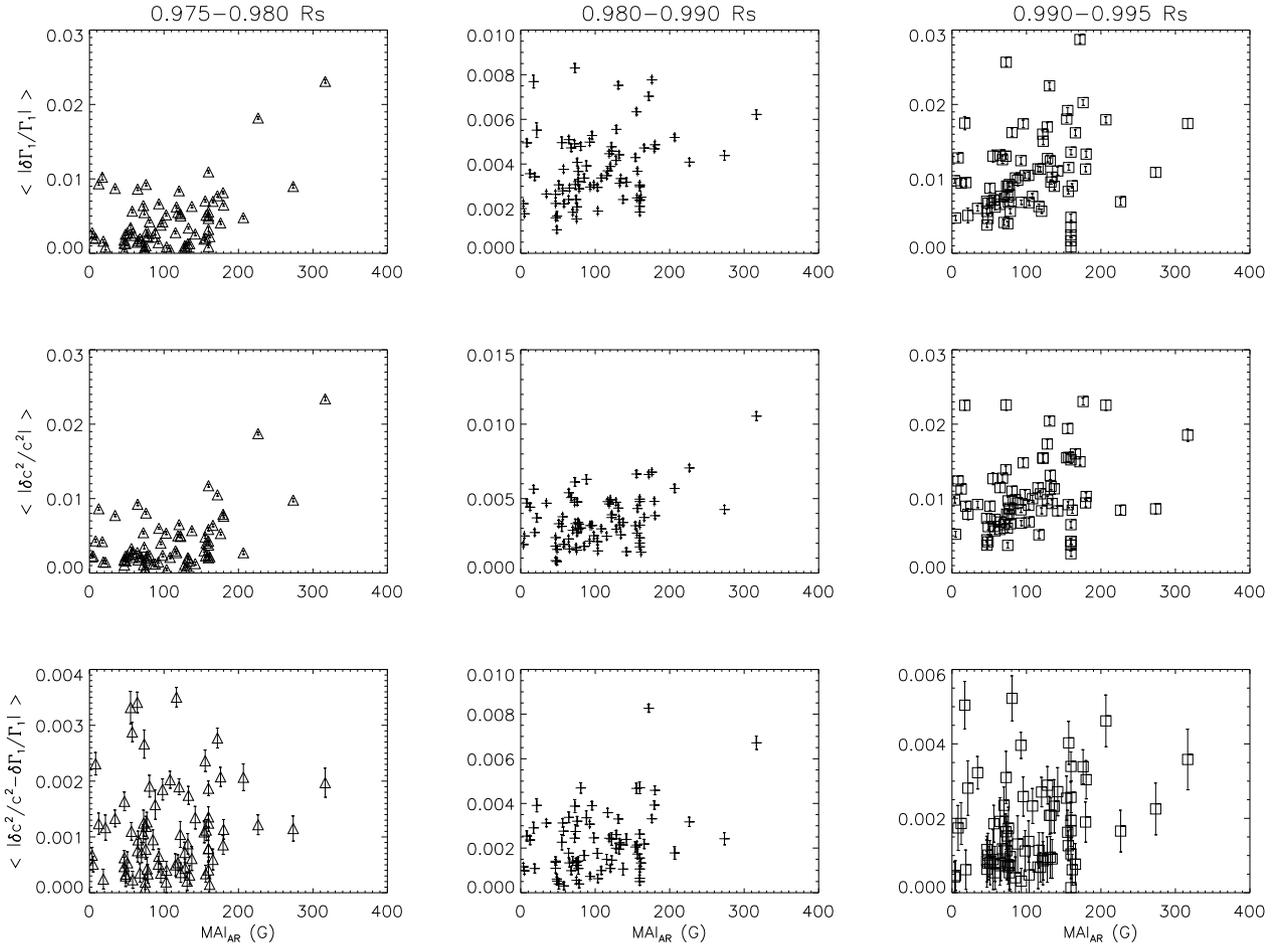}
\caption{
The scatter plots of
the subsurface structural disturbances in different depths vs. 
MAI of AR.
The arrangement of the panels is the same as in Fig.~\ref{fig:rdMAI}.
}
\label{fig:MAI}
\end{figure*}

\begin{figure*}
\small
\includegraphics[angle=90,width=17cm]{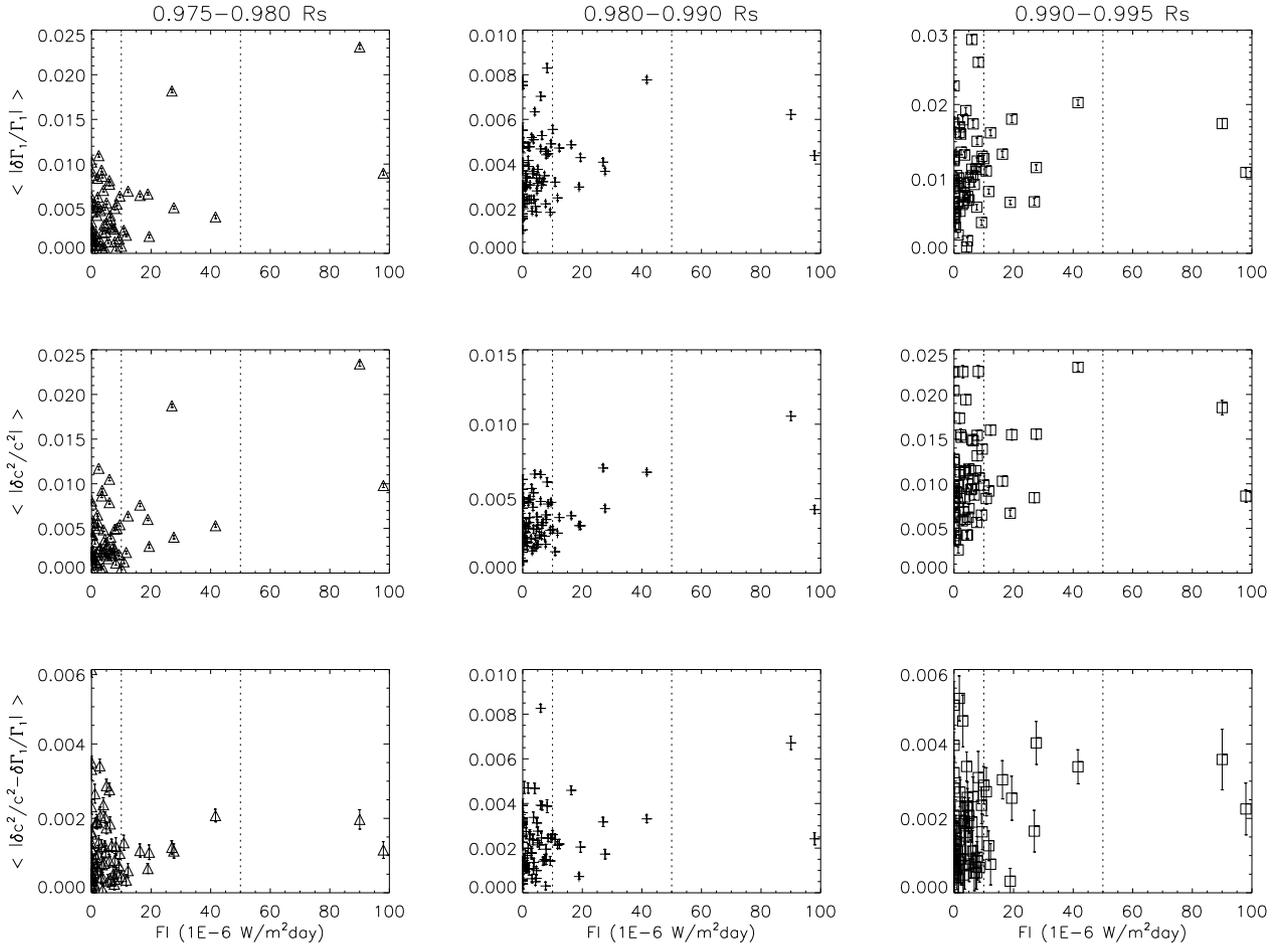}
\caption{
The scatter plots of
the subsurface structural disturbances in different depths vs. FI.
The arrangement of the panels is the same as in Fig.~\ref{fig:rdMAI}.
The dotted lines mark the locations of FI$=10$ and $50$,
within which a weak positive trend is visible in some plots 
(bottom left and middle middle).
}
\label{fig:FI100}
\end{figure*}

\begin{figure*}
\small
\includegraphics[angle=90,width=17cm]{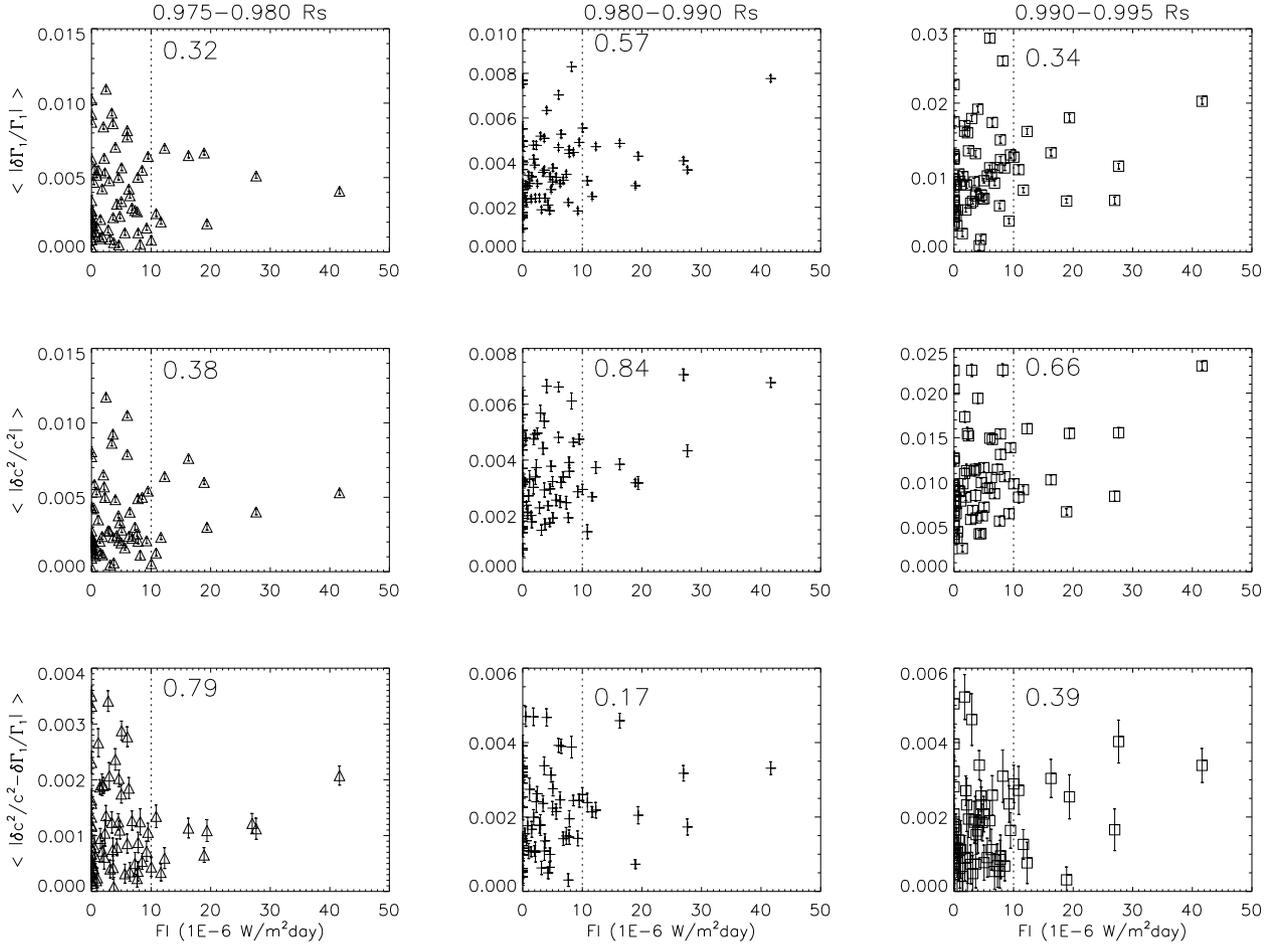}
\caption{
The scatter plots of
the subsurface structural disturbances in different depths vs. FI
for FI$<50$.
The arrangement of the panels is the same as in Fig.~\ref{fig:rdMAI}.
The dotted line marks the location of FI$=10$.
The number shown in each panel is the correlation coefficient
of the points in the range $10<$FI$<50$.
}
\label{fig:FI}
\end{figure*}

\begin{figure*}
\small
\includegraphics[angle=90,width=17cm]{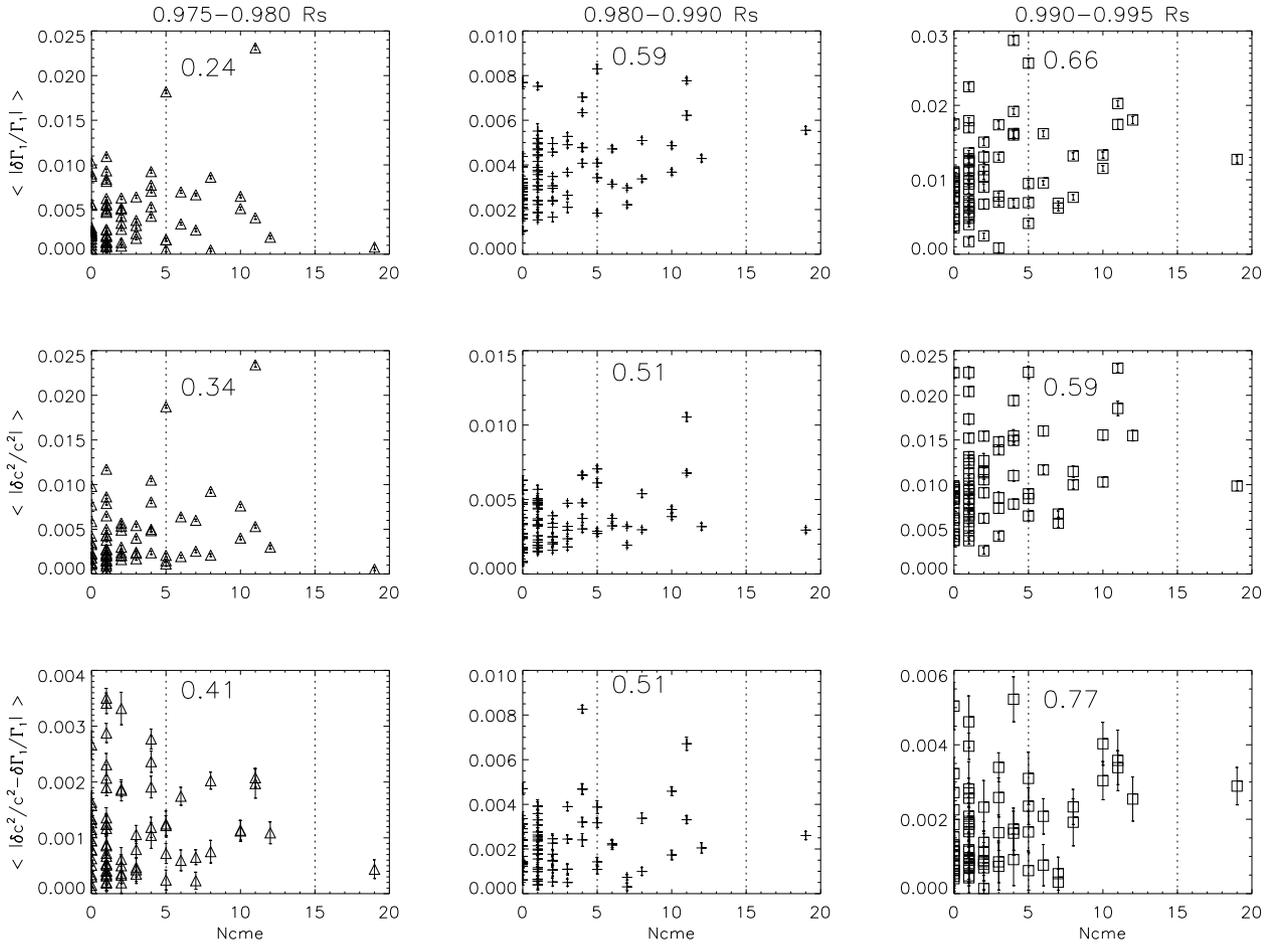}
\caption{
The scatter plots of
the subsurface structural disturbances in different depths vs. Ncme.
The arrangement of the panels is the same as in Fig.~\ref{fig:rdMAI}.
The dotted lines mark the locations of Ncme$=5$ and $15$,
within which the correlation coefficients are shown.
}
\label{fig:Ncme}
\end{figure*}

\begin{figure*}
\small
\includegraphics[angle=90,width=17cm]{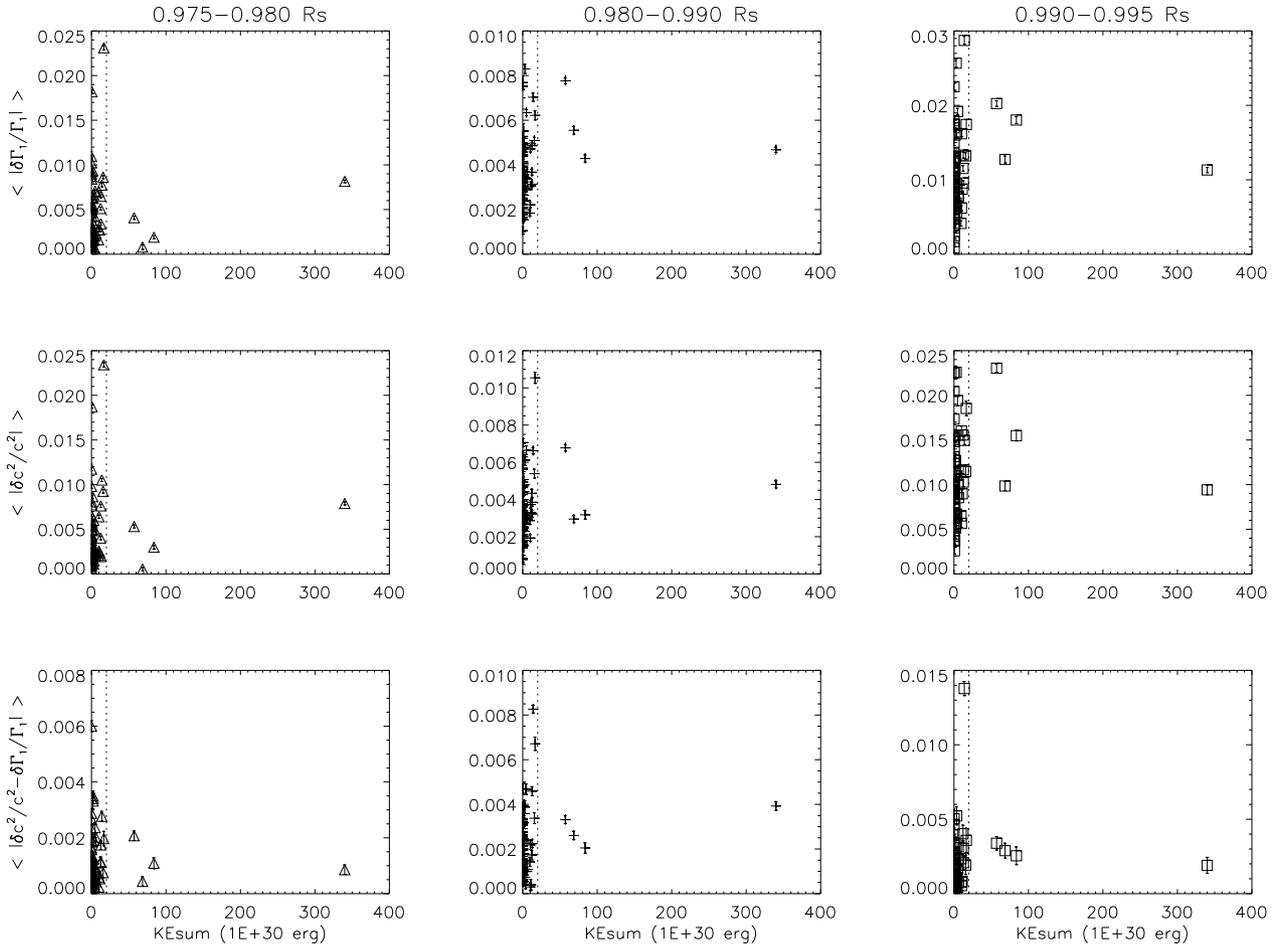}
\caption{
The scatter plots of
the subsurface structural disturbances in different depths vs. KEsum.
The arrangement of the panels is the same as in Fig.~\ref{fig:rdMAI}.
The dotted line marks the location KEsum $=20$,
below which the majority of the points are located.
}
\label{fig:KEsum400}
\end{figure*}

\begin{figure*}
\small
\includegraphics[angle=90,width=17cm]{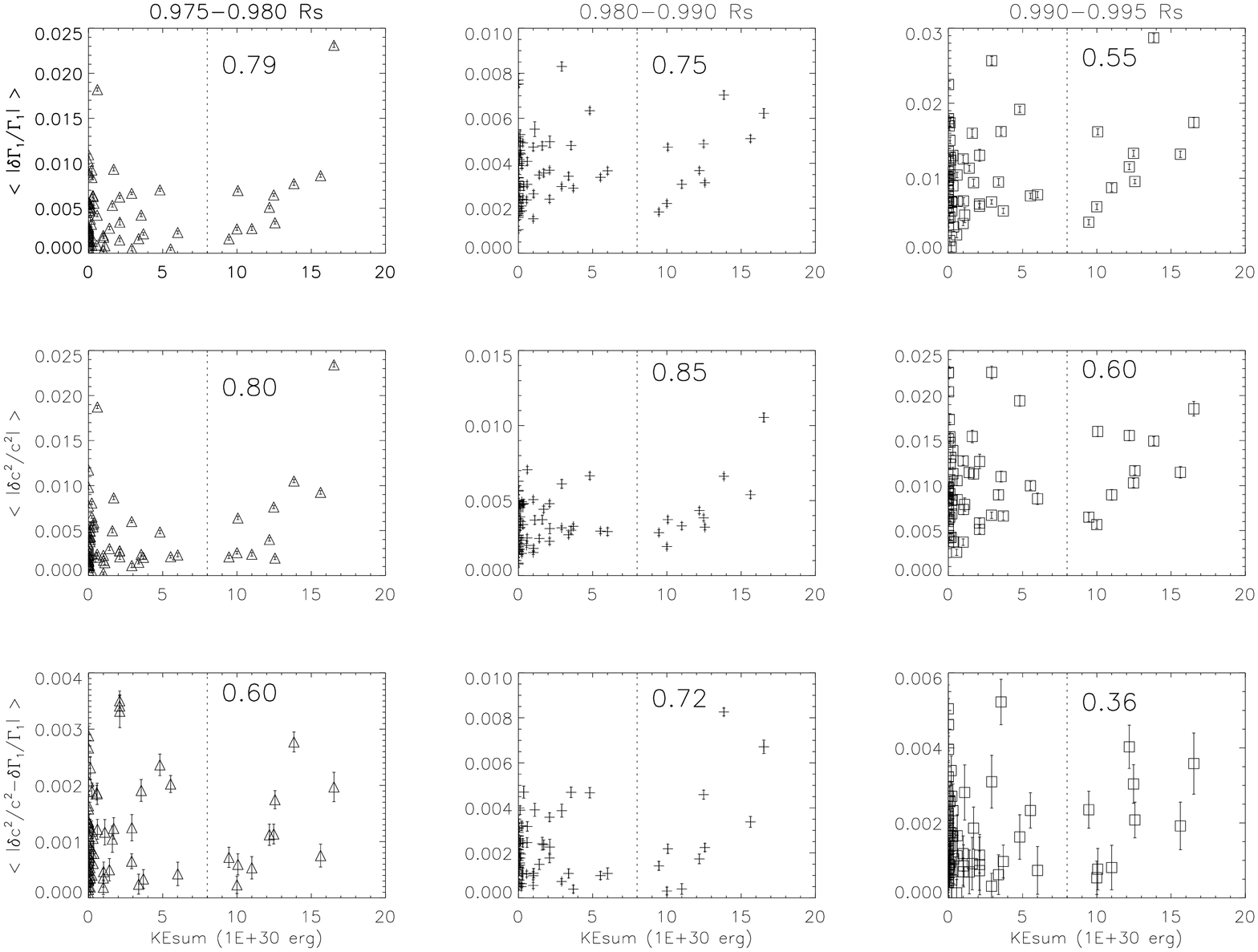}
\caption{
The scatter plots of
the subsurface structural disturbances in different depths vs. KEsum
for KEsum $<20$.
The arrangement of the panels is the same as in Fig.~\ref{fig:rdMAI}.
The dotted line marks the location KEsum $=8$.
A rising profile can be seen on the right of the line in most panels.
The number shown in each panel is the correlation coefficient calculated
for the points located within $8<$ KEsum $<20$.
}
\label{fig:KEsum}
\end{figure*}

The results of the subsurface relative differences versus 
rdMAI, dMAI, MAI$_{\rm AR}$, FI, Ncme and KEsum are plotted in 
Figures \ref{fig:rdMAI} to \ref{fig:KEsum}.
In each figure,
$<|\delta \Gamma_1/\Gamma_1|>$,
$<|\delta c^2/c^2|>$  and
$<|\delta c^2/c^2-\delta \Gamma_1/\Gamma_1|>$
are respectively placed in top, middle and bottom rows,
and  different depths are separated in different columns, 
as indicated on the top of each column.

Figure~\ref{fig:rdMAI} to \ref{fig:MAI} show
the results of Y(i,j) vs. different indices of the surface magnetic activity.
In all three figures, 
while
$<|\delta c^2/c^2|>$ shows a tighter distribution than
$<|\delta \Gamma_1/\Gamma_1|>$
in the layer $0.98-0.99R_\odot$,
the two becomes
almost identical in the deeper layer $0.975-0.98R_\odot$.
Figure~\ref{fig:rdMAI} shows no distinguishable regular pattern,
indicating that
the subsurface disturbances are uncorrelated with
the relative difference of MAI.
Interestingly,
the profiles in Figure~\ref{fig:dMAI} and \ref{fig:MAI} are almost identical,
suggesting that the relationships with dMAI and with MAI$_{\rm AR}$ are
very similar.
Therefore, the values of CC are only shown in Fig.~\ref{fig:dMAI}.
and the results of Fig.~\ref{fig:dMAI} and \ref{fig:MAI}
are discussed together in the following.
Most of the plots
do not show discernible regular patterns.
A weak positive trend is visible in some plots of
$<|\delta \Gamma_1/\Gamma_1|>$ and $<|\delta c^2/c^2|>$
(top two rows),
especially in the layer $0.975-0.98 R_\odot$.
The correlation coefficients of these plots, being only approximately $0.6$, 
are not sufficiently high to indicate a definitive correlation.
The earlier studies by \citet{bbb2013SoPh,Bogartetal2008SoPh}, however,
have claimed the existence of a positive correlation from their analysis.
It should be noted that
the subsurface anomalies analyzed in
the two earlier studies are the averages of {\em signed} relative differences.
The divisions of depth are also different between current and 
the two earlier studies.

The results of Y(i,j) vs. FI are plotted in
Figure~\ref{fig:FI100}.
The figure reveals that all but two points are located below FI $=50$ and 
that most points are concentrated in a small region of FI $<10$.
The two points with outstandingly high  FI are AR10488 and AR10656.
To better inspect the patterns in the populated region,
the region of FI $<50$ was re-drawn in Figure~\ref{fig:FI}.
The points below FI $=10$ in all plots of Figure~\ref{fig:FI}
are widely scattered
with no identifiable pattern.
In the region of FI $=10 - 50$, 
an approximately positive trend can be seen in a few panels.
The correlation coefficients for points in this range of FI
are shown in the corresponding panels.
There are three panels with CC higher than 0.6:
bottom left, 
$<|\delta c^2/c^2 - \delta \Gamma_1/\Gamma_1|>_{0.975-0.98 R_\odot}$,
middle middle, $<|\delta c^2/c^2|>_{0.98-0.99 R_\odot}$,
and middle right, $<|\delta c^2/c^2|>_{0.99-0.995 R_\odot}$.
Despite the high correlation coefficients,
there are only ten ARs in this FI range,
and the positive trend seems to appear in three arbitrary panels.
Therefore, no conclusion can be drawn regarding the
relationship between the flare productivity and 
the subsurface thermal disturbances.
The ten ARs with FI between 10 and 50 are listed in Table~\ref{tab:goodAR}.

The results of Y(i,j) vs. Ncme
are presented in Figure~\ref{fig:Ncme}.
Except for one outlying point at Ncme $=19$,
all other points are located below Ncme $\approx 13$.
The single exceptional point is AR09390.
Despite its high productivity of CMEs, 
the FI of AR09390 is only approximately 10, and
its subsurface thermal disturbances are no larger
than those of the rest of the data points.
Therefore,
in the figure of Y(i,j) vs. FI (Fig.~\ref{fig:FI}),
this point corresponds to a point 
at the lower left corner of the range in which
a positive trend is seen.
The distributions in Figure~\ref{fig:Ncme} in general 
show no clear regular pattern.
However,
in some panels in the middle and right columns of the figure,
a weak positive trend can be seen in the range Ncme $>5$.
The correlation coefficients for the points in this range (Ncme $=5-15$)
are printed in corresponding panels.
The values are mostly unremarkable,
suggesting that the productivity of CMEs is not strongly related
to most of the subsurface thermal properties.
The plot with the highest CC is
$<|\delta c^2/c^2 - \delta \Gamma_1/\Gamma_1|>$ 
in $0.99-0.995 R_\odot$
(CC$\approx 0.77$). 
While this suggests that certain mutual effects between 
CME production and $<|\delta c^2/c^2 - \delta \Gamma_1/\Gamma_1|>$
may be detectable just beneath the surface,
the number of ARs with Ncme larger than $5$ in our sample, being only twelve,
is insufficient to justify this implication.
The twelve ARs 
that form this positive trend are listed in Table~\ref{tab:goodAR}.

The results of Y(i,j) vs. KEsum
are shown in Fig.~\ref{fig:KEsum400}.
Most of the points are distributed below KEsum $\approx 20$
except for four points,
AR09433 (KEsum $=57.2$), 
AR09390 (KEsum $=68.5$), 
AR10792 (KEsum $=83.8$) and AR09787 (KEsum $=340$).
To better inspect the majority of points,
Y(i,j) vs. KEsum was re-plotted for the range of KEsum $\le 20$ 
in Figure~\ref{fig:KEsum}. 
The figure reveals a gap around KEsum $\approx 8$,
above which a positive trend is visible in all panels.
Below this gap, while the distribution patterns are more complex
than a linear trend, 
they are not as randomly and widely scattered as
the patterns in Fig.~\ref{fig:FI} and \ref{fig:Ncme}.
The correlation coefficients for the points located between 
KEsum $=8$ and $20$ are shown in corresponding panels.
Most of the values are equal to or higher than $0.6$,
indicating a good level of correlation.
However, this indication cannot be confidently verified in the current study
because this positive trend consists of only ten ARs.
These ten ARs are listed in Table~\ref{tab:goodAR}.
It is worth pointing out that
although the positive trends seen in the scatter plots of
the three eruptivity indices are all composed of approximately ten ARs,
there is less than 50\% overlap in the identity of the ARs,
as revealed in Table~\ref{tab:goodAR}.

\section{Summary} \label{sec:sum}

A statistical study of 77 ARs was conducted
to investigate
the possibility of correlations between 
the subsurface structural disturbances
of ARs and their surface magnetic activity
and coronal eruptivity.
The specific subsurface disturbances examined were
$<|\delta \Gamma_1/\Gamma_1|>$, $<|\delta c^2/c^2|>$
and $<|\delta c^2/c^2-\delta \Gamma_1/\Gamma_1|>$,
where $\Gamma_1$ and $c$ are two thermodynamic properties defined as
$\Gamma_1\equiv (\partial \ln P/\partial \ln \rho)_s$ and
$c \equiv \Gamma_1 P/\rho$,
in which $P$, $\rho$ and $s$ are pressure, density and entropy.
The absolute values
were averaged over three ranges of depth:
$0.975-0.98 R_\odot$, $0.98-0.99 R_\odot$ and $0.99-0.995 R_\odot$,
to represent the average structural disturbances in these layers.
The surface magnetic activity was measured by
the Magnetic Activity Index of AR (MAI$_{\rm AR}$) and
the relative and absolute differences of MAI between AR and QS (rdMAI, dMAI).
The coronal eruptivity level was gauged by
Flare Index (FI), total number of CMEs (Ncme) and total kinetic energy of
these CMEs (KEsum) of each AR.
The subsurface disturbances are denoted by Y(i,j) and
activity and eruptivity level indices by X(k) for simplicity.
The analysis consisted of visually inspecting the scatter plots
of different pairs of variables
and calculating the correlation coefficients
of the distribution patterns.

The subsurface anomalies and MAI$_{\rm AR}$ and dMAI
were reported to be positively correlated
by earlier studies \citep{Bogartetal2008SoPh,bbb2013SoPh}.
However,
the positive trend 
is only visible in the deepest layer $0.975-0.98 R_\odot$ in our analysis.
With a correlation coefficient value approximately $0.6$,
this correlation cannot be claimed by our study.
It should be noted that 
the quantities examined here are the averages of 
{\em unsigned} relative differences
while the two earlier studies had used the averages of {\em signed} values.
The divisions of the depth also differ between current and the earlier studies.
When Y(i,j) were plotted against
the relative MAI differences,
no correlation can be identified.

The scatter plots of Y(i,j) vs. three eruptivity indices indicate that
the distribution profiles change with
the magnitudes of the eruptivity indices.
In the region of low eruptivity,
the points are generally widely scattered with no distinguishable 
regular patterns.
The points only become more tightly and orderly distributed 
in the region with higher eruptivity,
and a positive trend with CC$\ge 0.6$ can be seen in some plots
(cf. Table~\ref{tab:good}).
Among the three eruptivity indices,
the positive trend is most prominent in KEsum.
The level of correlation is $\ge 0.6$ in most plots of Y(i,j) vs. KEsum.
In Y(i,j) vs. Ncme,
while the correlation level of the trend is in general low, 
a pattern of the correlation coefficient becoming higher in the shallower layer
can be seen.
The occurrence of the positive trend in Y(i,j) vs. FI, in contrast, 
does not show identifiable regularity.
These distributions with relatively high correlation levels
are all composed of only
$10-12$ points.
It is, therefore, uncertain whether 
the tight distribution
indicates that the correlations are detectable only
among sufficiently eruptive ARs
or after many strong eruptions,
as speculated in Sec.~\ref{sec:intro},
or whether it is simply a result of fewer points.
In addition,
many studies have pointed out that the helioseismic inversions based on
the identification of oscillation phases
can be contaminated by surface effects
\citep[e.g.][]{CR2007ApJ,Cally2009MNRAS}.
Although the subsurface structural differences here were obtained by
inverting the {\em frequency} differences
determined by
the ring-diagram analysis \citep{Hill1988ApJ}, 
the surface effects may distort the profile of the ring spectra
resulting in errors in the determination of frequencies
(Cally 2013, private discussion).
Therefore,
the ``positive trend'' may also partly be a result of
the inversion results containing effects propagating down
from the corona,
rather than an indication of a true correlation between 
the subsurface thermal structural disturbances and the coronal eruptivity.
Therefore,
no definitive correlations can be claimed at this stage.
To verify the correlations suggested by the analysis in this study,
we will need to first
improve the current Helioseismic inversion procedures,
and then apply the structural inversions to more ARs with high eruptivity. 
A theoretical study of the relationship
between the subsurface disturbances and the eruptivity in the atmosphere
is also necessary.
Lastly,
it is interesting to note that
the distribution pattern seems to depend little on the magnitude of
the subsurface disturbances.
In other words,
considering only the ARs with larger, or smaller, subsurface disturbances
does not lead to more ordered distribution.

 \begin{table*}
 \small
 \caption{List of the ARs following a linear trend in the scatter plots of
different eruptivity indices}
  \label{tab:goodAR}
 \begin{tabular}{|l|l|}
 \tableline  
Y(i,j) vs. FI & 09390, 09433, 09782, 09893, 09899, 09901, 09906, 09907,
10792, 10875 \\ 
  &  \\
Y(i,j) vs. Ncme & 
08545, 09402, 09433, 09461, 09893, 09899, 09901, 09948, 10656, 10792, 10875
\\ 
  & \\
Y(i,j) vs. KEsum  & 08534, 08545, 09402, 09893, 09896, 09901, 09934, 09948,
 10656 10875 \\
  & \\
 \tableline 
 \end{tabular}
 \end{table*}

 \begin{table*}
 \small
 \caption{List of the plots showing a trend with correlation coefficient
higher than $0.6$ (CC$>0.6$)}
  \label{tab:good}
 \begin{tabular}{|c|cccc|}
 \tableline  
 &  dMAI and MAI$_{\rm AR}$ & $10<$FI$<50$ & $5<$Ncme$<15$ & $8<$KEsum$<20$ \\ \hline
                             & 
                              $0.975-0.98 R_\odot$ &  
                              No correlation &        
                                             &        
                              $0.975-0.98 R_\odot$\\   
$|\delta \Gamma_1/\Gamma_1|$ &  
                                                   &  
                                                 &    
                                                 &    
                            $0.98-0.99 R_\odot$ \\  
                           & 
                                                & 
                                                & 
                            $0.99-0.995 R_\odot$ & 
                                         \\ \hline 
                   & 
                    $0.975-0.98 R_\odot$ &  
                                         &   
                    No correlation       &   
                    $0.975-0.98 R_\odot$ \\  
$|\delta c^2/c^2|$ & 
                                         &  
                    $0.98-0.99 R_\odot$  &   
                                         &   
                    $0.98-0.99 R_\odot$ \\  
                  & 
                                         &  
                    $0.99-0.995 R_\odot$  &   
                                         &   
                    $0.99-0.995 R_\odot$ \\ \hline 
                                           & 
                                              No correlation &  
                                              $0.975-0.98 R_\odot$ & 
                                                                   &  
                                              $0.975-0.98 R_\odot$ \\ 
$|\delta c^2/c^2-\delta \Gamma_1/\Gamma_1|$ & 
                                                            &  
                                                            & 
                                                                   &  
                                              $0.98-0.99 R_\odot$ \\ 
                                            & 
                                                            &  
                                                            & 
                                            $0.99-0.995 R_\odot$   &  
                                               \\ 
 \tableline 
 \end{tabular}
 \end{table*}

%

%
\acknowledgments
This work is funded by
the NSC of ROC under grant NSC99-2112-M-008-019-MY3
and the MOE grant ``Aim for the Top University'' to 
the National Central University. 
The author wish to thank Dean-Yi Chou for helpful inputs and suggestion.

%
\bibliographystyle{spr-mp-nameyear-cnd}  
\bibliography{ref}                

\end{document}